\documentclass[a4paper]{jpconf}

\usepackage{graphicx}

\begin{document}

\title{Spectroscopy of the heaviest nuclei (theory)}

\author{A.\ V.\ Afanasjev and H. Abusara}
\address{Department of Physics and Astronomy, Mississippi State
University, Mississippi State, Mississippi 39762, USA}

\author{E.\ Litvinova}
\address{GSI Helmholtzzentrum f\"{u}r Schwerionenforschung,
D-64291 Darmstadt, Germany}

\address{Institut f\"{u}r Theoretische Physik,
Goethe-Universit\"{a}t, D-60438 Frankfurt am Main, Germany}

\author{P.\ Ring}
\address{Physik-Department der Technischen Universit\"{a}t M\"{u}nchen, D-85748 Garching,
Germany}

\ead{afansjev@erc.msstate.edu}

\begin{abstract}
Recent progress in the applications of covariant density functional
theory (CDFT) to the description of the spectroscopy of the heaviest
nuclei is reviewed. The analysis of quasiparticle spectra in
actinides and the heaviest $A\sim 250$ nuclei provides a measure of
the accuracy of the description of single-particle energies in CDFT
and an additional constraint for the choice of effective interactions
for the description of superheavy nuclei. The response of these
nuclei to the rotation is rather well described by cranked
relativistic Hartree+Bogoliubov theory and it serves as a
supplementary tool in configuration assignment in odd-mass nuclei. A
systematic analysis of the fission barriers with allowance for
triaxial deformation shows that covariant density functional theory
is able to describe fission barriers on a level of accuracy
comparable with the best phenomenological macroscopic+microscopic
approaches.
\end{abstract}


\section{Introduction}


The study of the nuclei around $^{254}$No, the heaviest elements for
which detailed spectroscopic data can be obtained with the current
generation of facilities \cite{HG.08}, is intimately connected with
the search for superheavy nuclei \cite{HG.08,AKF.03}. This is because
spectroscopic information such as the energies of single-particle
states and the rotational response can be used to constrain
theoretical models and their parameterizations. The need for such a
constraint is dictated  by the fact that different theoretical models
predict different locations for the ``island of stability'' of
shell-stabilized superheavy nuclei. Macroscopic+microscopic methods
based on different phenomenological potentials localize this island
around the $Z=114$ and $N=184$ spherical shell closures
\cite{Nilsson69,CDHMN.96,MN.94}. Self-consistent calculations based
on the Skyrme energy density functionals (EDF) show this island
around the $Z=126$ and $N=184$ spherical shell closures for most of
the forces \cite{CDHMN.96,BRRMG.99}. Covariant  density functional
theory (CDFT) \cite{VALR.05} localizes this island around $Z=120$ and
$N=172$ \cite{AKF.03,BRRMG.99}.

\begin{figure}[h]
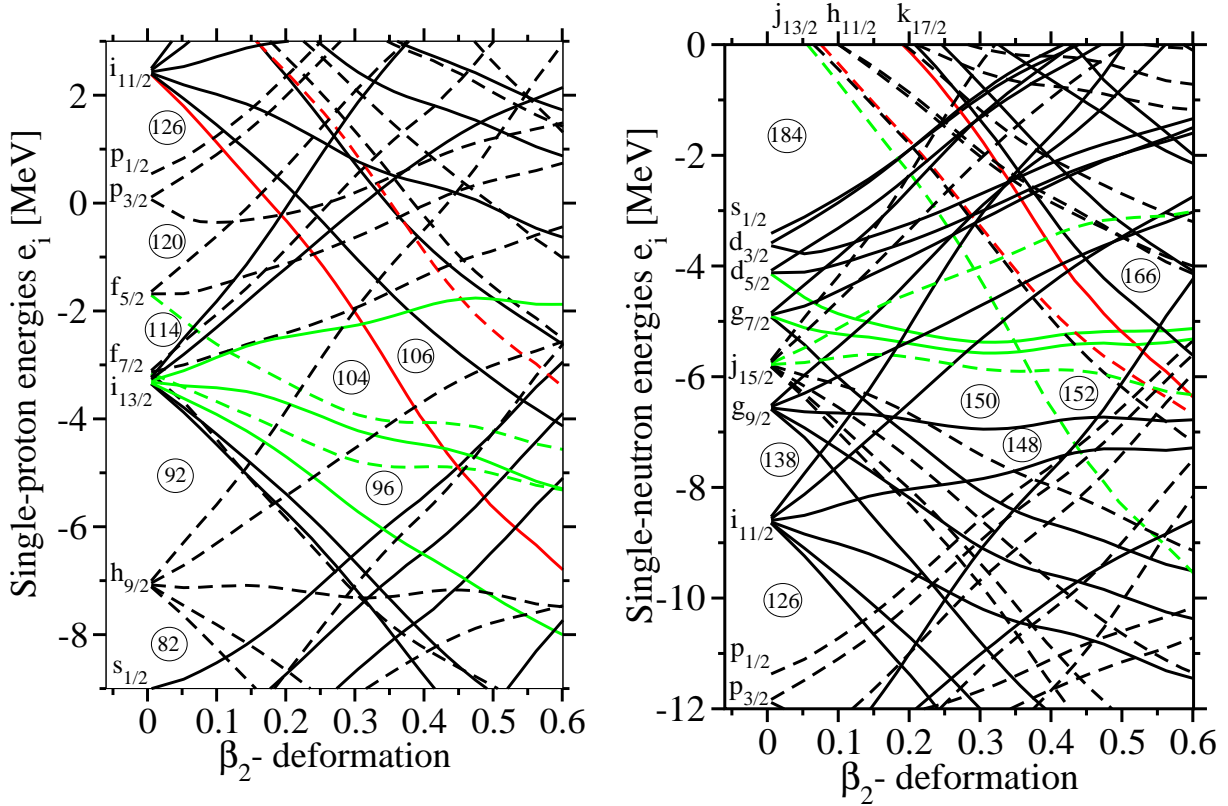

\begin{minipage}{18pc}
\includegraphics[width=18pc]{Nilsson-p.eps}
\end{minipage}\hspace{0.5cm}%
\begin{minipage}{18.5pc}
\includegraphics[width=18.5pc]{Nilsson-n.eps}
\end{minipage}
\begin{minipage}{37.5pc}
\caption{\label{Nilsson} Proton and neutron single-particle energies
in $^{252}$No as a function of the quadrupole deformation $\beta_2$.
They are obtained in RMF+BCS calculations using the constant gap
approximation and the NL1 parameterization of the RMF Lagrangian.
Solid and dashed lines are used for positive and negative parity
states, respectively. Green lines are used for the single-particle
orbitals observed in odd-mass nuclei around $^{250}$Fm, while red
lines are used for the $\Omega=1/2$ orbitals that may be observed in
nuclides with $N \approx 162$  and/or $Z \approx 108$ \cite{AKF.03}.}
\end{minipage}
\end{figure}

 Apart of the question of the location of the ``island of stability'',
the question of the stability of superheavy nuclei is another puzzle
defining the physics of superheavy nuclei. The probability for the
formation of a superheavy nucleus in a heavy-ion-fusion reaction and
its survival are directly connected to the height of its fission
barrier~\cite{IOZ.02}. Thus, the experimental data on fission
barriers in the actinide region provide an important testing ground
for density functional theories (DFT).

It is impossible to review all efforts undertaken in the study of
theoretical aspects of the spectroscopy of the heaviest nuclei; thus
we will concentrate on the results obtained in the framework of
covariant density functional theory. The manuscript is organized in
the following way. In Sec.\ \ref{sp-energies}, the accuracy of the
description of the energies of deformed one-quasiparticle states in
the actinide region and its consequences for spherical shell closures
in superheavy nuclei are analyzed. The rotational response in
odd-mass nuclei and its applicability for configuration assignment
will be discussed in Sec.\ \ref{sect-rot}. The description of inner
fission barriers in CDFT is discussed in Sec.\ \ref{fission-sect}.
Finally, Sec.\ \ref{sect-concl} summarizes our main conclusions.

\begin{figure}[h]
\includegraphics[width=18pc]{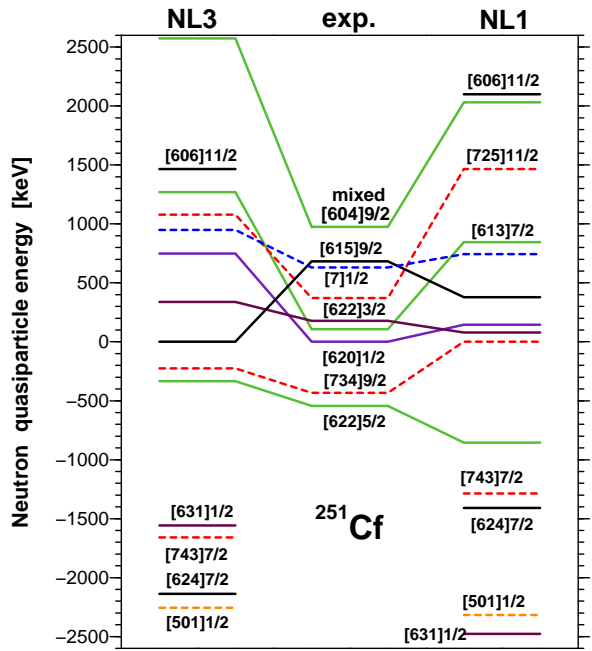}\hspace{2pc}%
\begin{minipage}[b]{18pc}\caption{\label{fig-Cf251}
Experimental and theoretical one-quasiparticle energies of neutron
states in $^{251}$Cf. Positive and negative energies are used for
particle and hole states, respectively. The experimental data are
taken from Refs.\ \protect\cite{251Cf,251Cf-old,251Cf-a}, while the
results of calculations from Ref.\ \cite{AKF.03}. Solid and dashed
lines are used for positive and negative parity states, respectively.
The symbols 'NL3' and 'NL1' indicate the RMF parameterization. Colors
indicate the spherical subshells from which deformed
one-quasiparticle states emerge, namely, green - $2g_{9/2}$, black -
$1i_{11/2}$, indigo - $3d_{5/2}$, maroon - $2g_{7/2}$, red -
$1j_{15/2}$ and blue - $1j_{13/2}$.}
\end{minipage}
\end{figure}

\section{Single-particle degrees of freedom}
\label{sp-energies}

One way of obtaining information on the energies of spherical
subshells active in the vicinity of expected shell closures in the
``island of stability'' is by studying well-deformed
one-quasiparticle states in heavy actinide and/or light
transfermium nuclei. As illustrated in the Nilsson diagram of Fig.\
\ref{Nilsson}, the deformation causes the spherical single-particle
states to split and forces low-$\Omega$ states emerging from high-$j$
spherical subshells (for example, the $\pi f_{5/2}$ and $\pi
i_{11/2}$ subshells, see left panel of Fig.\ \ref{Nilsson}) to come
closer to the Fermi level in much lighter systems.

Nuclei around $^{254}$No represent the end of the region in which
reliable experimental data on the structure of deformed
one-quasiparticle states are available \cite{HG.08}. Although the
decay studies of heavier odd-mass nuclei such as $^{265}$Hs and
$^{259}$Sg (see, for example, Ref.\ \cite{Decay} and references
therein) probe one-quasiparticle states, these nuclei cannot be used
for a meaningful comparison with theory because of the tentative
nature of the configuration assignment. This is even more true for
odd-mass nuclei with $Z\geq 105$ which are identified by only a few
events.

The first ever fully self-consistent description of deformed
one-quasiparticle states in the framework of CDFT was presented in
Ref.\ \cite{AKF.03} on the example of the $^{249,251}$Cf and
$^{249}$Bk nuclei. More systematic calculations covering the
rare-earth and actinide regions and a number of the CDFT
parameterizations are currently in progress \cite{SA.10}. Since these
results confirm the general observations obtained in Ref.\
\cite{AKF.03}, we will illustrate the main features of the
description of quasiparticle spectra of deformed nuclei on the
example of  $^{251}$Cf (see Fig.\ \ref{fig-Cf251}).

Two types of discrepancies between theory and experiment are visible
in Fig.\ \ref{fig-Cf251}. {\it First, the calculated spectra are less
dense than experimental ones.} This is especially visible in the
pairs of deformed states emerging from the same spherical subshells
(namely, the $\nu 9/2[734]$ and $\nu 11/2[725]$ states from the $\nu
1j_{15/2}$ subshell as well as the $\nu 5/2[622]$ and $\nu 7/2[613]$
states from the $\nu 2g_{9/2}$ subshell), the energy gaps between
which are considerably stretched as compared with experiment. {\it
Second, while the calculated energies of a number of states are
rather close to experiment, the energies of some states and their
relative positions deviate substantially from experiment.}  For
example, only NL1 gives the correct ground state $\nu 1/2[620]$ in
$^{251}$Cf, whereas NL3 gives the $\nu [615]9/2$ (Fig.\
\ref{fig-Cf251}). The displacement of the calculated centroids of the
above discussed pairs of deformed states from experimental ones
clearly suggests that the discrepancies between experiment and
calculations can be traced back to energies of spherical subshells
from which deformed states emerge.

\begin{figure*}
\begin{center}
\includegraphics[width=12.0cm]{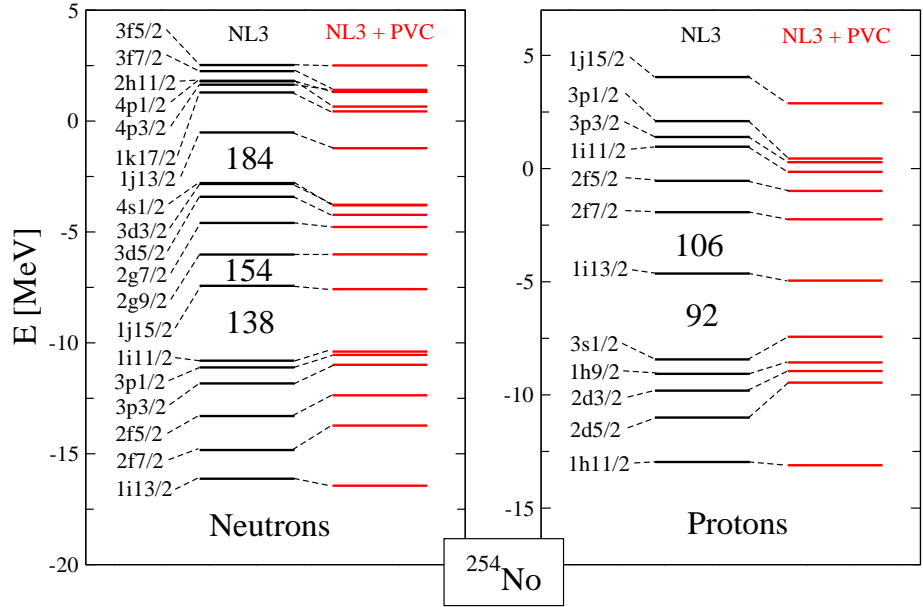}
\end{center}
\vspace{0.3cm} \caption{\label{fig-241Am} Neutron and proton
single-particle states in $^{254}$No $(Z=102,N=152)$. The left column
in each panel shows the spectra obtained in pure RMF calculations,
while right column the spectra computed within RMF with allowance for
the particle-vibration coupling. The calculations are performed at
spherical shape employing the NL3 parameterization (from Ref.\
\cite{LA.10}).}
\label{No254}
\end{figure*}

Comparing theory and experiment it was concluded in Ref.\
\cite{AKF.03} that in the NL1 and NL3 parameterizations, the energies
of the spherical subshells, from which the deformed states in the
vicinity of the Fermi level of the $A\sim 250$ nuclei emerge, are
described with an accuracy better than 0.5 MeV for most of the
subshells. The discrepancies (in the range of $0.6-1.0$ MeV) are
larger for the $\pi 1h_{9/2}$ (NL3, NL1), $\nu 1i_{11/2}$ (NL3), $\nu
1j_{15/2}$ (NL1) and $\nu 2 g_{9/2}$ (NL3) spherical subshells.
Considering that the RMF parameterizations were fitted only to bulk
properties of spherical nuclei this level of agreement is good. In
contrast, the accuracy of the description of single-particle states
is unsatisfactory in the NLSH and NL-RA1 parameterizations. This
clearly indicates that before extrapolating to superheavy nuclei the
quality of the parameterization with respect of the description of
single-particle energies has to be tested.

These results clearly indicate the need for an improvement in the
description of the single-particle energies within CDFT. There are a
number of possibilities which need further exploration. The
stretching of the energy scale in CDFT calculations as compared with
experiment is related to the low effective mass (Lorentz mass in the
notation of Ref.\ \cite{JM.89}) $m^*(k_F)/m \approx 0.66$ of the
nucleons at the Fermi surface in CDFT. It has been demonstrated for
spherical nuclei that particle-vibration coupling brings the average
level density in closer agreement with experiment in relativistic
\cite{LR.06} and non-relativistic calculations \cite{MBBD.85}. This
effect is clearly visible in Fig.\ \ref{No254} where the
particle-vibration coupling leads to a pronounced increase of the
level density around the Fermi surface both for proton and neutron
subsystems comparatively to the pure RMF spectra. Similar effects are
expected in deformed nuclei. However, the corrections to the energies
of quasiparticle states in odd nuclei due to particle-vibration
coupling are expected to be less state dependent in deformed nuclei
because (due to fragmentation) the surface vibrations are less
collective in deformed nuclei than in spherical ones \cite{S.76}.

The measured and calculated energies of the single-particle states at
normal deformation provide constraints on the spherical shell gaps of
superheavy nuclei. Such an analysis restricts the choice of CDFT
parameterizations only to those which predict $Z=120$ and $N=172$ as
shell closures in superheavy nuclei \cite{AKF.03}. The inclusion of
particle-vibration coupling compresses the single-particle spectra in
the $(Z=120, N=172)$ nucleus and decreases the size of above
mentioned gaps. However, this nucleus still remains a doubly magic
spherical superheavy nucleus in CDFT \cite{LA.10}.

\section{Rotational degrees of freedom}
\label{sect-rot}

Additonal information on the structure of single-particle states can
be obtained by studying the rotational response in odd-mass nuclei.
This is especially important for nuclei at the edge of the region
where spectroscopic studies are possible (the nuclei with masses
$A\sim 255$), for which alternative methods of configuration
assignment are not feasible. To illustrate that we use rotational
structures in $^{241}$Am. The rotational bands based on the Nilsson
orbitals $\pi 5/2[642]$ (from the $i_{13/2}$ subshell), $\pi
5/2[523]$ (from the $h_{9/2}$ subshell) and $\pi 3/2[521]$ (from the
$f_{7/2}$ subshell) have been observed in this nucleus in Ref.\
\cite{Am241}. As can be seen in the left panel of Fig.\
\ref{fig-241Am}, at low frequencies they have distinctly different
kinematic moments of inertia $J^{(1)}$. The theoretical
interpretation of these bands has been performed in cranked
relativistic Hartree+Bogoliubov (CRHB) theory \cite{ARK.00,AKF.03}
employing the NL1 parameterization of the RMF Lagrangian and the D1S
Gogny force in the pairing channel. In addition, approximate particle
number projection has been carried out by means of the Lipkin-Nogami
(LN) method. Theoretical calculations (right panel of Fig.\
\ref{fig-241Am}) describe well the absolute values of the kinematic
moment of inertia of different configurations and their evolution
with rotational frequency. In particular, the splitting of two
signatures of the $\pi 5/2[642]$ configuration is rather well
described in the model calculations. On the contrary, the $\pi
5/2[523]$ and $\pi 3/2[521]$ bands show (with exception of very low
frequencies in the case of the $\pi 3/2[521]$ band) no signature 
splitting. Model calculations for the two signatures of the 
$\pi 5/2[523]$ configuration show explicitly this
feature. Unfortunately, it was not possible to get convergence in the
case of the $\pi 3/2[521]^+$ \footnote{The superscript to the orbital
label is used to indicate the sign of the signature $r$ for that
orbital $(r=\pm i)$.} configuration. However, the analysis of the
quasiparticle routhian diagram confirms that the $\pi 3/2[521]^\pm$
configurations have to be degenerate in energy up to rotational
frequency $\Omega_x \sim 0.16$ MeV in agreement with experimental
observations. At higher frequencies, small signature separation is
expected in the calculations.

In addition to the above mentioned features, the relative properties
of different bands both with respect of each other and with respect
to $^{240}$Pu (which has one proton less) are well described in the
model calculations. The increase of the kinematic moment of inertia
in the bands of $^{241}$Am as compared with the ground state band in
$^{240}$Pu is caused by the blocking effect which results in a
decreased proton pairing.

\begin{figure*}
\begin{center}
\includegraphics[width=12.0cm]{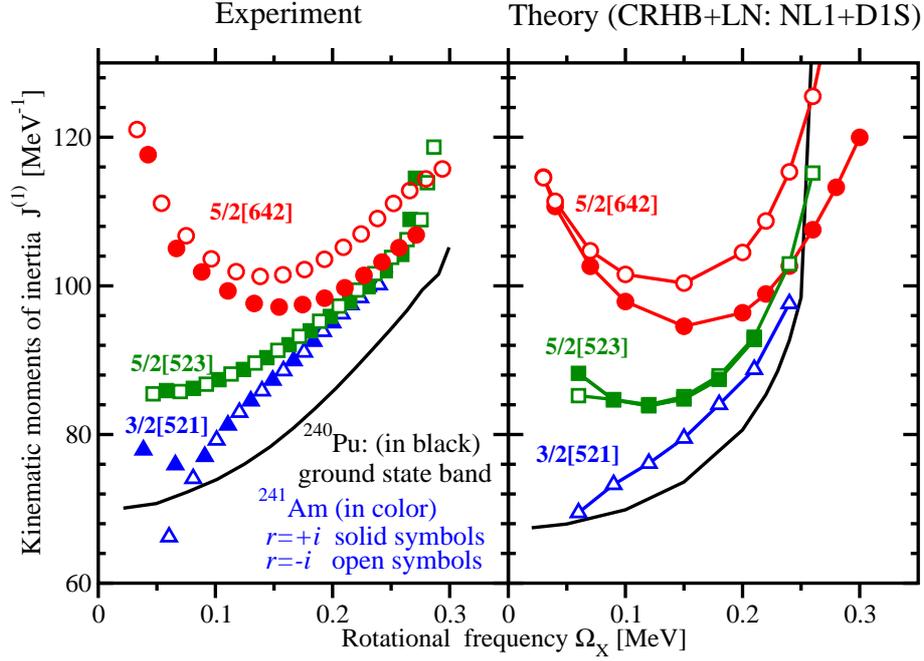}
\end{center}
\caption{\label{fig-241Am} Experimental and calculated kinematic
moments of inertia $J^{(1)}$ of the indicated one-quasiproton
configurations in $^{241}$Am. In addition, these moments are shown
for the ground state band in $^{240}$Pu. Experimental data are taken
from Refs.\ \cite{Am241} ($^{241}$Am) and \cite{Pu240} ($^{240}$Pu).}
\label{pot}
\end{figure*}

\begin{figure}[h]
\includegraphics[width=21pc]{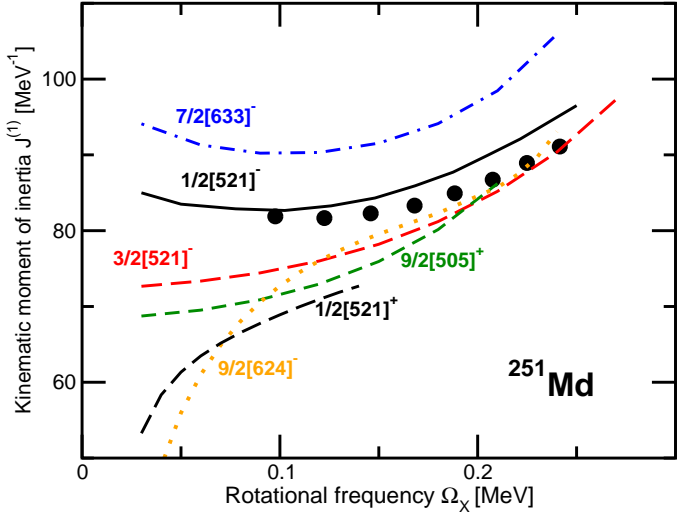}\hspace{2pc}%
\begin{minipage}[b]{14pc}
\caption{\label{fig-Md251} {Calculated kinematic moments of inertia
$J^{(1)}$ of the indicated one-quasiproton configurations in
$^{251}$Md are shown by lines. The configurations $7/2[633]$,
$9/2[505]$, $9/2[624]$, $3/2[521]$ are either signature degenerate
or characterized by small signature splitting. Thus, for simplicity
we show the results only for one signature.  The kinematic moments of
inertia for observed band \cite{Md251} are shown by solid circles.}}
\end{minipage}
\end{figure}

Fig.\ \ref{fig-Md251} shows another example of the description of
rotational bands in CRHB calculations. The single decoupled band
observed recently in the odd-proton nucleus $^{251}$Md \cite{Md251}
has been assigned to the $\pi 1/2[521]^{-}$ configuration. CRHB
calculations have been performed for several low-lying
configurations. However, the $\pi 7/2[633]$, $\pi 3/2[521]$, $\pi
9/2[624]$ and $\pi 9/2[505]$ configurations can be excluded from
consideration because they lead either to signature degenerate bands
or to bands with small signature splitting; as a consequence, both
signatures are expected to be observed in experiment. On the other
hand, the calculations for the $\pi 1/2[521]^-$ configuration
describe the experimental data rather well thus confirming the
configuration assignment given in Ref.\ \cite{Md251}. Note that the
results obtained with the NL3 parameterization of the RMF Lagrangian
are close to the ones obtained with NL1. Thus, they do not alter the
interpretation of the observed band.

 These two examples outline the current approach to the configuration
assignment of one-quasiparticle states in odd-mass nuclei based on their 
rotational response. They clearly show that the rotational properties 
reflected through
\begin{itemize}
\item the absolute values of the kinematic moment of inertia (especially at low
rotational frequencies) and their evolution with rotational frequency

\item the presence or absence of signature splitting

\item the relative properties of different configurations with respect of each
other and/or with respect to the neighboring even-even nucleus
\end{itemize}

provide a useful tool for single-particle configuration
assignment. However, it is necessary to recognize that this method of
configuration assignment has to be complemented by other independent
methods and has to rely on sufficient experimental data. The
interpretation of the rotational band in $^{253}$No is quite
illustrative in this respect. Initially, it was interpreted as based
on the $\nu 7/2[624]$ configuration \cite{No253-1}. However, improved
experiments allowed to identify the M1 transitions between opposite
signatures of the observed band \cite{No253-1} which led to the $\nu
9/2[734]$ configuration assignment. The kinematic moments of inertia
of the observed band under two configuration assignments are
described within a typical theoretical uncertainty\footnote{The
calculations show that the kinematic moments of inertia in the nuclei
of interest are described with an accuracy better than 10\% of their
absolute value, see current results and the results presented in
Refs.\ \cite{AKF.03,Fm250}. The only exception is the rotational band in $^{255}$Lr, where the
discrepancy of around 15\% is seen under the $\pi [624]9/2$
configuration assignment \cite{Lr255}. On the basis of the moment of
inertia, CRHB calculations favor the $\pi 7/2[633]$ assignment. Note
that the kinematic moments of inertia for the assigned  $\pi
7/2[633]$ and $\pi 9/2[624]$ rotational bands in $^{251}$Es are
described in CRHB calculations within 7\% accuracy \cite{Lr255}.},
and, as a result, the configuration assignment based only on
rotational properties cannot be reliable. The branching ratios of
observed M1 and E2 transitions have to be used in order to
distinguish different configuration assignments \cite{No253-2}.


\section{Fission barriers: the role of triaxiality}

\label{fission-sect}


The (static) inner fission barriers $B_{f}^{st}$ are important for
several physical phenomena. The size of the fission barrier is a
measure for the stability of a nucleus (which decays by spontaneous
fission) reflected in its spontaneous fission lifetime~\cite{SP.07}.
The probability for the formation of a superheavy nucleus in a
heavy-ion-fusion reaction is also directly connected to the height of
its fission barrier~\cite{IOZ.02}. The height $B_{f}^{st}$ is a
decisive quantity in the competition between neutron evaporation and
fission of a compound nucleus in the process of its cooling
\cite{IOZ.02}. The population and survival of hyperdeformed states at
high spin also depends on the fission barriers~\cite{DPS.04,AA.08}.
In addition, the $r-$process of stellar nucleosynthesis depends
(among other quantities such as masses and $\beta$-decay rates) on
the fission barriers of very neutron-rich nuclei~\cite{AT.99,MPR.01}.

However, the progress in the study of the fission barriers within
covariant density functional theory has been slower than in its
non-relativistic counterparts \cite{KALR.10}. Until now the majority
of the calculations have been performed under the restriction to
axial symmetry (see Ref.\ \cite{KALR.10} and references therein).
However, axially symmetric calculations cannot be directly compared
with experimental data since, as has been shown earlier in
non-relativistic calculations (see Refs.\ \cite{WERP.02,SBDN.09} and
references therein), the lowering of fission barriers due to
triaxiality is significant and can reach $3-4$ MeV in some nuclei. So
far, the impact of triaxiality on the height of the inner fission
barrier has only been studied in specific nuclei such as $^{264}$Hs
\cite{BRRMG.98} and $^{240}$Pu \cite{BHR.03} within the RMF+BCS
approach as well as $^{240}$Pu \cite{LNV.10} within the RHB approach.

In order to fill this gap in our knowledge, a systematic
investigation of the inner fission barriers within the triaxial
RMF+BCS approach has been performed for the first time \cite{AAR.10}
employing the recently developed NL3* parameterization \cite{NL3*} of
the RMF Lagrangian. The truncation of the basis is performed in such
a way that all states belonging to the shells up to $N_{F}=20$
fermionic shells and $N_{B}=20$ bosonic shells are taken into
account. In the pairing channel, we use a seniority pairing force,
with the strength parameters defined as in Ref.\ \cite{AAR.10}:
\begin{eqnarray}
A \cdot G_n = 9.1 - 6.4 \frac{N-Z}{A}\,\,\,\,{\rm MeV} \\
A \cdot G_p = 8.1 + 10.0 \frac{N-Z}{A}\,\,\,\,{\rm MeV}
\label{G-strength}
\end{eqnarray}
Note that all states in the pairing window $E_k<E_{\rm cutoff}=120$
MeV are taken into account. Because the nuclei considered are all
well bound, pairing is treated in the BCS approximation.

The calculations are performed imposing constraints on the axial and
triaxial mass quadrupole moments. The method of quadratic constraints
uses a variation of the function
\begin{eqnarray}
\left<H\right> + \sum_{\mu=0,2} C_{2\mu}
(\langle\hat{Q}_{2\mu}\rangle-q_{2\mu})^2
\end{eqnarray}
where $\left<H\right>$ is the total energy, and
$\langle\hat{Q}_{2\mu}\rangle$ denotes the expectation values of the
mass quadrupole operators
\begin{eqnarray}
\hat{Q}_{20}&=&2z^2-x^2-y^2 \\
\hat{Q}_{22}&=&x^2-y^2
\end{eqnarray}
In these equations, $q_{2\mu}$ is the constrained value of the
multipole moment, and $C_{2\mu}$ the corresponding stiffness
constants \cite{RS.80}.

\begin{figure*}
\includegraphics[width=16.0cm]{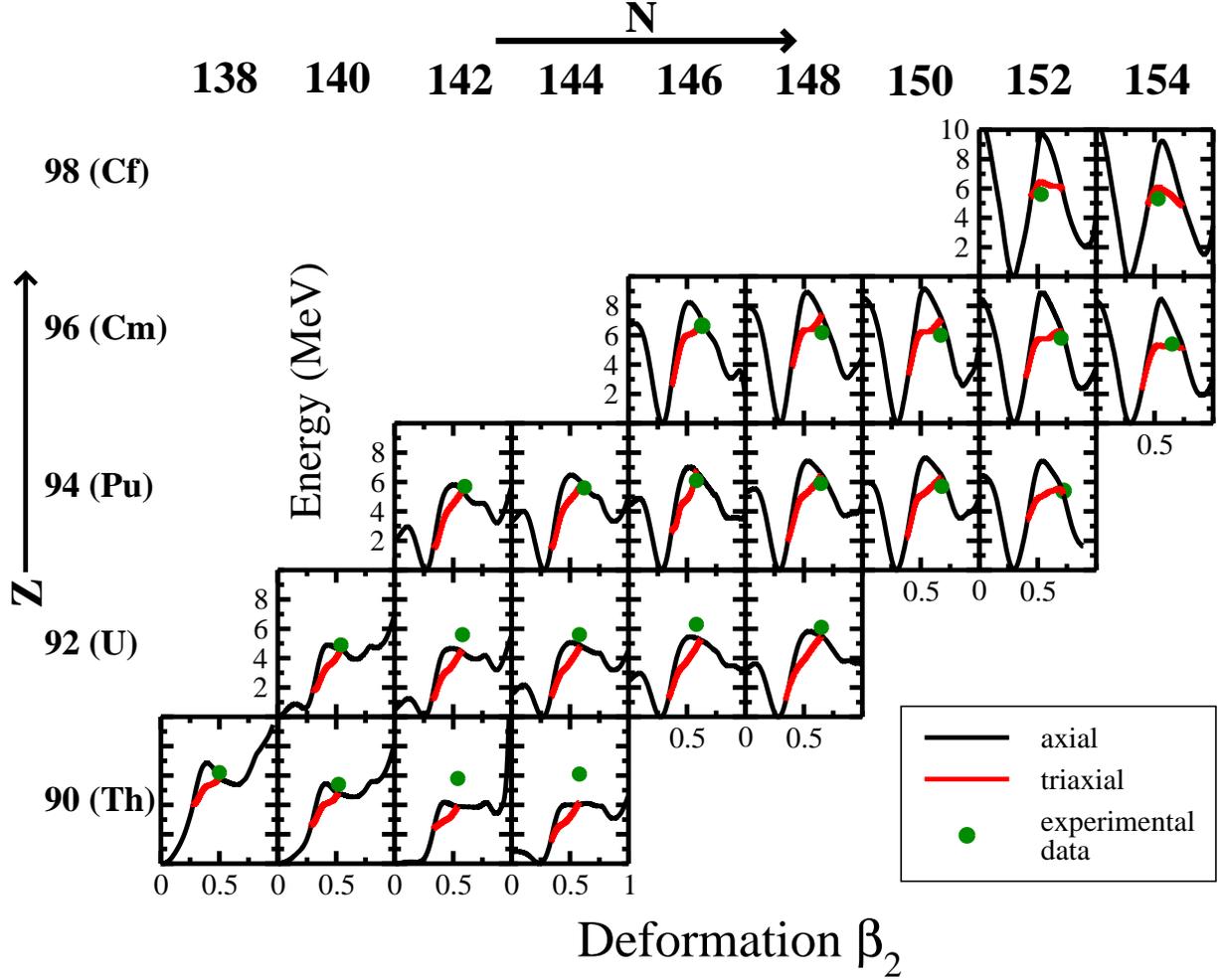}
\caption{Deformation energy curves of even-even actinide nuclei
obtained in RMF+BCS calculations with the NL3* parameterization.
Experimental data are taken from Table IV in Ref.~\cite{SGP.05}. A
typical uncertainty in the experimental values, as suggested by the
differences among various compilations, is of the order of $\pm0.5$
MeV~\cite{SGP.05}. See text for detail.}
\label{pot}
\end{figure*}

The results of these systematic calculations are shown in Fig.\
\ref{pot}. The deformation energy curves (shown by full black lines
in Fig.\ \ref{pot}) for axially symmetric solutions are obtained as
the $\gamma=0^{\circ}$ cross-section of the potential energy
surfaces. The deformation energy curves for triaxial solutions are
obtained by the minimization of the potential energy surfaces along
the $\beta_2$-direction. The deformation energy curves along the
triaxial fission path are shown by red full curves. We show
deformation energy curves for triaxial solutions only in the range of
the $\beta_2$ values at which it is lower in energy than the
deformation energy curve of the axially symmetric solution. Note that
the potential energy surfaces are normalized to zero energy at the
normal-deformed minimum. One can see that by allowing for triaxiality
the fission barrier heights are reduced by $1-3$ MeV as compared with
axially symmetric solutions. This lowering depends on the proton and
neutron numbers. It also brings on average the results of
calculations in closer agreement with experimental data shown by
green solid circles in Fig.\ \ref{pot}. These circles display the
height of the experimental fission barrier at the calculated
$\beta$-deformation of the saddle point. On average calculated
$\gamma$-deformations of the triaxial parts of the fission path are
close to 10$^{\circ}$.

\begin{figure}[h]
\includegraphics[width=14pc]{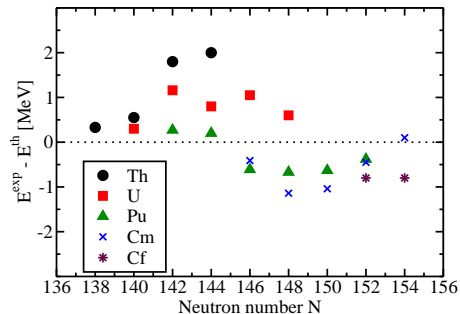}\hspace{2pc}%
\begin{minipage}[b]{19pc}%
\caption{\label{Th-vs-exp-N} The difference between experimental and
calculated heights of inner fission barriers as a function of neutron
number $N$.}
\end{minipage}
\end{figure}

Fig.\ \ref{Th-vs-exp-N}  shows the differences between calculated and
experimental heights of inner fission barriers. The average deviation
between theory and experiment is 0.76 MeV. This is comparable with
the results obtained in the macroscopic+microscopic method (see Sec.
IVC and Fig. 11 in Ref.\ \cite{DPB.07} and Sec. VII A in Ref.\
\cite{MSI.09} which describe experimental fission barriers with an
average error of around 1 MeV. It is necessary, however, to say that
neither the proton nor the neutron particle number dependences of
fission barrier height are completely reproduced in the calculations.
This is exemplified in Fig.\ \ref{Th-vs-exp-N}. However, the same
problem exists also in macroscopic+microscopic calculations (see
Fig.\ 11 in Ref.\ \cite{DPB.07} and Figs.\ 23-32 in Ref.\
\cite{MSI.09}). There are very few energy density functional
calculations of the fission barriers which include triaxial
deformations, and neither of them confronts in a systematic way
experimental data in actinides. However, the limited results obtained
with the Skyrme energy density functionals presented in Ref.\
\cite{BQS.04} show similar unresolved particle number dependences for
the inner fission barrier heights.

\section{Conclusions}
\label{sect-concl}

The analysis of quasiparticle spectra in actinides and heaviest $A\sim 250$ 
nuclei provides a  measure of the accuracy of the description of single-particle 
energies in CDFT. In addition,  it restricts the choice of the CDFT parameterizations 
to only those which predict $Z=120$ and $N=172$ as shell closures in spherical
superheavy nuclei. The inclusion of particle-vibration coupling does not change the 
position of shell closures for these nuclei.

The description of rotational properties in odd-mass nuclei within
the framework of the cranked relativistic Hartree+Bogoliubov
framework has been discussed. It is shown that rotational properties
are reasonably well described. As a consequence, their theoretical
analysis provides a useful supplementary tool in the assignment of
single-particle configurations to rotational bands observed at the
edge of the region where spectroscopic studies are possible.

The first systematic investigation of fission barriers in the
actinide region with triaxiality accounted has been performed within
covariant density functional theory. It is found that with only one
exception ($^{234}$Th) in all the nuclei under investigation the
height of the inner fission barrier is reduced by allowing for
triaxial deformations by $1-3$ MeV. A systematic comparison of our
results with experimentally determined fission barriers in this
region shows reasonable agreement with data comparable with the best
macroscopic+microscopic calculations.

\section{Acknowledgments}

This work has been supported by the U.S. Department of Energy under
the grant DE-FG02-07ER41459 and by the DFG cluster of excellence
\textquotedblleft Origin and Structure of the Universe\textquotedblright\ 
(www.universe-cluster.de). E. L. is supported by the Hessian LOEWE initiative 
through the Helmholtz International Center for FAIR.

\section*{References}

\end{document}